\documentclass[twocolumn,showpacs,preprintnumbers,amsmath,amssymb,prb]{revtex4}
\usepackage{graphicx}
\usepackage{dcolumn}
\usepackage{bm}
\usepackage{float}

\begin{document}

\title{Proposal to determine the Fermi-surface topology of a doped iron-based
superconductor using bulk-sensitive Fourier-transform Compton scattering}

\author{Yung Jui Wang$^1$, Hsin Lin$^1$,
B. Barbiellini$^1$, P.E. Mijnarends$^{1,2}$, S. Kaprzyk$^{1,3}$, R.S.
Markiewicz$^1$ and A. Bansil$^1$}

\affiliation{
$^1$Physics Department, Northeastern University, Boston, Massachusetts
02115, USA\\
$^2$Department of Radiation, Radionuclides and Reactors, Faculty of
Applied Sciences, Delft University of Technology, Delft, The
Netherlands \\
$^3$AGH University of Science and Technology, 30059 Krakow, Poland\\
}

\date{\today}

\begin{abstract}
We have carried out first-principles calculations of the Compton
scattering spectra to demonstrate that the filling of the hole Fermi
surface in LaO$_{1-x}$F$_{x}$FeAs produces a distinct signature in the
Fourier transformed Compton spectrum when the momentum transfer vector
lies along the [100] direction. We thus show how the critical
concentration $x_c$, where hole Fermi surface pieces are filled up and the
superconductivity mediated by antiferromagnetic spin fluctuations is
expected to be suppressed, can be obtained in a bulk-sensitive manner.
\end{abstract}


\maketitle
\def\thesection{\arabic{section}}


The Fermi surface (FS) topology is a key ingredient for high temperature
superconductivity in iron based layered pnictides. The so-called $s_{\pm}$
model \cite{mazin,mazin2,chubukov,chubukov2} predicts superconducting gaps
of one sign on the FS cylindrical hole sheets near $\Gamma(0,0)$ and of
another sign on cylindrical electron sheets at M$(\pi,\pi)$. Doping $x$ is needed
to move the system away from the magnetic instabilities due to FS nesting
\cite{singh,mazin3,norman}. In the superconducting material, spin
fluctuations (related to residual FS nesting) may provide a glue for the
Cooper pairs \cite{mazin}. However, a complete filling of the hole FS at a
certain electron doping will eventually lead to the suppression of the
spin fluctuation glue. The exact value $x_c$ of this critical doping might
be affected by subtle correlations effects \cite{haule}.

Experimental information regarding the FS topology comes mostly from
angle resolved photoemission spectroscopy (ARPES)
which is a surface sensitive probe.\cite{norman,Terashima,dhlu,Sekiba,Sahrakorpi}
. Moreover, since the doping level in the bulk could be
different from that at the surface \cite{mazin2}, the FS signal should be
checked with bulk probes. Since the FS information from quantum
oscillation studies\cite{AIC,jga} can be distorted because of the required
high magnetic fields, we suggest determination of the FS topology via
Compton scattering measurements. \cite{footnote_posi,ft_posi1}
In this letter, we show that a one dimensional Fourier transform of
the Compton profile along [100] presents a large signal when the hole
Fermi surface vanishes, providing a bulk sensitive method for determining
the critical doping $x_c$ for high temperature
superconductivity in LaO$_{1-x}$F$_{x}$FeAs.

Recent advances in synchrotron light sources and detector
technology have renewed interest in high-resolution Compton scattering as
a bulk probe of fermiology related issues, see e.g.,
Refs.\cite{Tanaka,Huotari,Stutz,cooper}.
In a Compton scattering experiment, one measures a
directional Compton profile (CP), $J(p_z)$, which is related to the twice
integrated ground-state electron momentum density $\rho(p_x,p_y,p_z)$ by
\begin{equation}
\label{eq1}
J(p_z) = \int \int \rho(p_x,p_y,p_z) dp_xdp_y~,
\end{equation}
for high momentum and energy transfer\cite{Kaplan}. The exploration of FS
topology with the aid of Compton scattering is complicated by the double
integral in Eq. \ref{eq1}. As a result, FS breaks in $\rho(p_x,p_y,p_z)$
do not usually induce rapid variations in $J(p_z)$. A possible approach to
deal with this problem is to measure CPs along many different directions
and use state-of-the-art reconstruction methods based on the
autocorrelation function $B(x,y,z)$ to obtain $\rho(p_x,p_y,p_z)$
\cite{matsumoto}. A much simpler and more robust method will be proposed
in this letter.


The calculations presented here were performed within the local density
approximation (LDA) using an all-electron fully charge self-consistent
semi-relativistic Korringa-Kohn-Rostoker (KKR) method \cite{ABkkr}. The
compound LaO$_{1-x}$F$_x$FeAs has a simple tetragonal structure
(space-group P4/nmm). We have used the experimental lattice parameters
\cite{Qiu2008} of LaO$_{0.87}$F$_{0.13}$FeAs in which no spin-density-wave
order was observed in neutron scattering experiments. Self-consistency was
obtained for $x=0$ and the effects of doping $x$ were treated within a
rigid band model by shifting the Fermi energy to accommodate the proper
number of electrons \cite{footnote_sophi,ft_sophi1,ft_sophi2,ft_sophi3,ft_sophi4}.
The convergence of the crystal potential was
approximately $10^{-4}$ Ry.
The electron momentum density (EMD)
$\rho(p_x,p_y,p_z)$ was computed on a fine mesh of $40.4\times10^{6}$
within a sphere of radius $12$ a.u. in momentum space. To simulate the
effect of resolution in high resolution Compton scattering experiments,
the EMD has been been convoluted with a Gaussian characterized by a FWHM
of $0.17$ a.u.

\begin{figure}
\includegraphics[width=6.0cm]{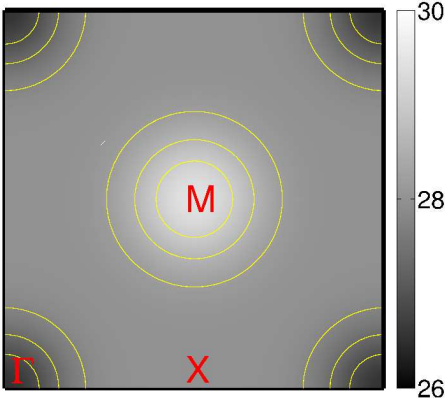}
\caption{(Color online)
Calculated LCW distribution of in the
paramagnetic tetragonal Brillouin zone
of LaOFeAs.}
\label{lcw}
\end{figure}

Since the LaO$_{1-x}$F$_{x}$FeAs electronic structure has a
two-dimensional character, we shall focus on the calculated (001)
2D-projection of the momentum density given by
\begin{equation}
\rho^{2D}(p_x,p_y) = \int \rho(p_x,p_y,p_z) dp_z~.
\label{eq2}
\end{equation}
In experiments, one adopts the so-called direct Fourier-transform method
\cite{matsumoto} to reconstruct $\rho^{2D}(p_x,p_y)$ from several
directional CPs $J(p_z)$ measured in the (001) plane. This method
uses the autocorrelation function $B(x,y,z)$ which is
straightforwardly defined as the Fourier transformation of the
momentum density
\begin{eqnarray}
 B(x,y,z)&=&\int \int \int dp_xdp_ydp_z~ \\ \nonumber
    & \times &  \rho(p_x,p_y,p_z)\exp[i(p_{x}x+p_{y}y+p_{z}z)]~.
\end{eqnarray}
Since $\rho(p_x,p_y,p_z)$ can be expressed as a sum over the momentum
density of the natural orbitals $\psi_{j}(x,y,z)$ \cite{bba01} by using
the convolution theorem, it can be shown that $B(x,y,z)$ is the
autocorrelation of the natural orbitals
 \begin{eqnarray}
B(x,y,z)&=&\sum_{j}n_j\int \int \int dudvdw \\ \nonumber
   &\times&\psi_{j}(x+u,y+v,z+w)\psi_{j}^{*}(u,v,w)~,
 \end{eqnarray}
\begin{figure}
\includegraphics[width=7cm]{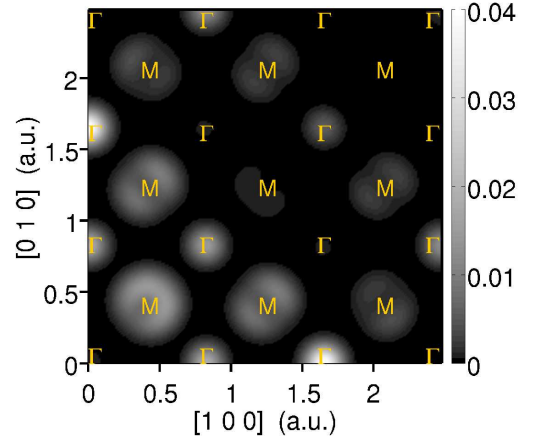}
\caption{(Color online)
$\Delta \rho|_{x_1}^{x_2}(p_x,p_y)$
for $x_2=0.15$ and $x_1=0.10$.
The yellow labels indicate the high symmetry positions $\Gamma$ and M in
momentum space.
}
\label{fseh_a}
\end{figure}
\begin{figure}
\includegraphics[width=7cm]{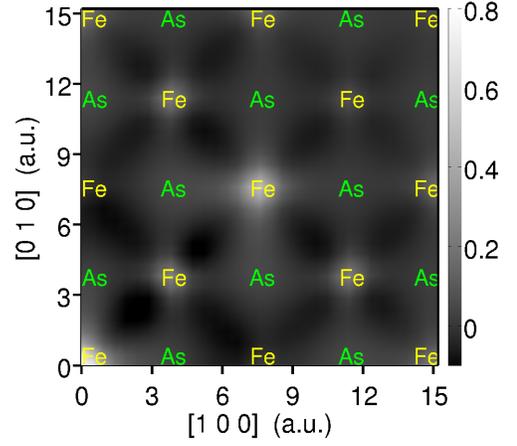}
\caption{(Color online)
$\Delta B|_{x_1}^{x_2}(x,y)$ for $x_2=0.15$ and $x_1=0.10$.
The yellow labels Fe indicate the atoms in the iron sublattice;
green labels As indicate the $(x,y)$-projection of the atoms
in the arsenic sublattice. Atomic assignments are based on
assumption Fe is at origin.
}
\label{fseh_b}
\end{figure}
where $n_j$ is the occupation number of the natural orbital
$\psi_{j}(x,y,z)$. In the experiments $B(x,y,z)$ is obtained directly
along a given direction by taking the 1D-Fourier transform of the CP along
that direction. Then, once a set of $B$'s has been calculated, a fine mesh
is set up in real space and $B(x,y,z)$ is obtained at every mesh point by
interpolation. Finally, if desired, an inverse Fourier transform of
$B(x,y,0)$ yields the distribution $\rho^{2D}(p_x,p_y)$.
Our simulations reveal that the breaks in $\rho^{2D}(p_x,p_y)$ caused by FS
crossings are scattered throughout momentum space with small weights given
by matrix elements involving mostly the Fe $d$ orbitals. Therefore, FS
features are not easily detected directly in the $\rho^{2D}(p_x,p_y)$
distribution. However, as shown in Fig.~\ref{lcw} the Lock-Crisp-West (LCW)
folding \cite{lcw1973} can
enhance FS breaks by coherently superposing the umklapp terms according to
\begin{equation}
n(k_x,k_y)=\sum_{G_x,G_y}\rho^{2d}(k_x+G_x,k_y+G_y)~,
\end{equation}
where $n(k_x,k_y)$ gives the number of occupied states at the point
$(k_x,k_y)$ in the first Brillouin zone by summing over all projected
reciprocal lattice vectors $(G_x,G_y)$. The maximum of $n(k_x,k_y)$
at $M(\pi,\pi)$ is associated with the electron pockets while
the minimum at $\Gamma$ ($0$,$0$) is related to the hole pockets.
Since the LCW folding can also enlarge artificial errors from the
experimental data,
below we will introduce a more robust means of extracting information
about the evolution of the FS topology with doping.

\begin{figure}
\includegraphics[width=7cm]{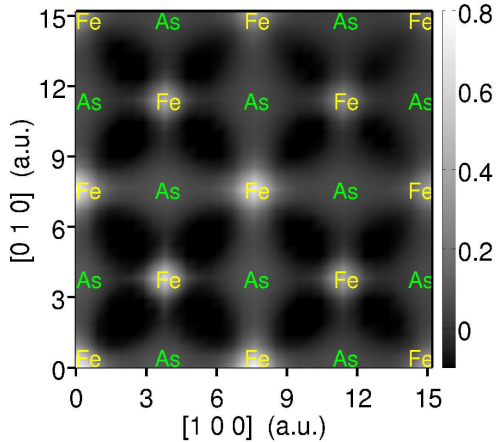}
\caption{(Color online)
$\Delta B|_{x_1}^{x_2}(x,y)$. for $x_2=0.15$ and $x_1=0.10$,
after artificially removing all the Bloch states around M.
Atomic assignments are based on assumption Fe is at origin.
}
\label{fsh_10-15_b}
\end{figure}

We can get more precise information on wave function symmetry near the
Fermi surface (FS) by taking difference maps between two nearby
dopings,
\begin{equation}
\Delta \rho|_{x_1}^{x_2}(p_x,p_y)=\rho^{2D}(p_x,p_y)|_{x_2}-
\rho^{2D}(p_x,p_y)|_{x_1}~,
\label{eq6}
\end{equation}
where $x_2$ and $x_1$ are two different doping levels such that $x_2>x_1$.
The subtraction in Eq.~\ref{eq6} acts as a projector on the Fermi level
subspace with the advantage of eliminating the large isotropic
contribution of the core and some irrelevant valence electrons.
The difference for $x_2=0.15$ and $x_1=0.10$
shown in Fig.~\ref{fseh_a}
displays interesting FS effects
strongly modulated by Fe $d$
wave function effects.
The corresponding Fourier transform $\Delta B|_{x_1}^{x_2}(x,y)$,
Fig.~\ref{fseh_b}, separates the different length scales in real space,
which contribute to the oscillations in $\Delta
\rho|_{x_1}^{x_2}(p_x,p_y)$.
Thus, the peaks in the autocorrelation function $\Delta B$ indicate
characteristic distances over which wave functions at the Fermi level
are coherent. The peaks in Fig.~\ref{fseh_b} mostly stem from the Fe $d$
orbitals since these largely dominate at the Fermi level
\cite{singh}. In fact, from Fig.~\ref{fseh_b} one can see that the
main peaks correlate very well with the iron sublattice.
However, note that there are weaker features, marked "As",
which correlate with the positions of the As atoms.
Since the bands near the Fermi level mostly consist of Fe
$d_{xz}$, $d_{yz}$ and $d_{x^2-y^2}$ orbitals\cite{weiku}
, Fig.~\ref{fseh_b} reveals these characters in real space.
To facilitate comparisons we normalize $\Delta B$ to unity at the origin.

\begin{figure}
\includegraphics[width=7cm]{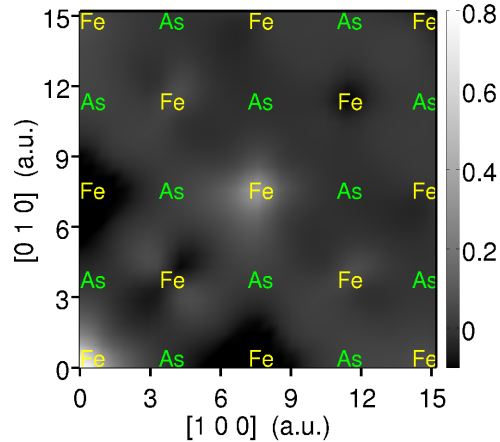}
\caption{(Color online)
$\Delta B|_{x_1}^{x_2}(x,y)$. for $x_2=0.15$ and $x_1=0.10$,
after artificially removing all the Bloch states around $\Gamma$.
Atomic assignments are based on assumption Fe is at origin.
}
\label{fse_10-15_b}
\end{figure}

We employ filtering techniques to enhance the sensitivity of the
$\Delta B$-maps to particular FS cylinders. Thus, in Fig.~\ref{fsh_10-15_b}
we artificially remove the FS around M by applying a filter cutting out the
Bloch states near M. As a result some Fe peaks essentially disappear,
revealing the wavefunction characters of the $\Gamma$ cylinders.  In the
same way, Fig.~\ref{fse_10-15_b} shows the corresponding maps for the
electron cylinders, generated by filtering out the cylinders at $\Gamma$ .
This filtering procedure can be tested by calculating $\Delta
B|_{x_1}^{x_2}(x,y)$ at higher
doping  for $x_2$ and $x_1$, where the number of added electrons
is sufficient to remove the hole cylinders at $\Gamma$ without any
filtering needed.\cite{Sekiba,mazin3,haule,footnote1}
The result is very similar to Fig.~\ref{fse_10-15_b},
confirming that the present filter is an efficient way of sorting out
contributions from different FS cylinders\cite{footnote_filter}.
By comparing Fig.~\ref{fsh_10-15_b}
and Fig.~\ref{fse_10-15_b}, it is clear that the hole FSs and the electron
FSs give strikingly different contributions to $\Delta B|_{x_1}^{x_2}(x,y)$.

In Fig.~\ref{fsh_10-15_b} [hole FSs], there is a peak at every
Fe site, with a corresponding weaker network of peaks at As sites.  In
contrast, for the electron contribution in Fig.~\ref{fse_10-15_b}, there are
dips at some Fe sites and the As sites signatures are not visible.
Some of these features can be readily understood with the phase factor
$\exp(i{\bf k\cdot r})$ of the Bloch wave function.
At the band bottom,
$\Gamma$, ${\bf k}=0$ and all atoms are in phase ({\it bonding}).  At the
top of the band, $(\pi ,\pi )$, the phase factor is $(-1)^{m+n}$ -- i.e.,
perfectly {\it antibonding}.  Since the FSs are small cylinders near
$\Gamma$ and $(\pi ,\pi )$ respectively, their $\Delta B$-maps are dominated
by this interference term.

\begin{figure}
\includegraphics[width=8cm]{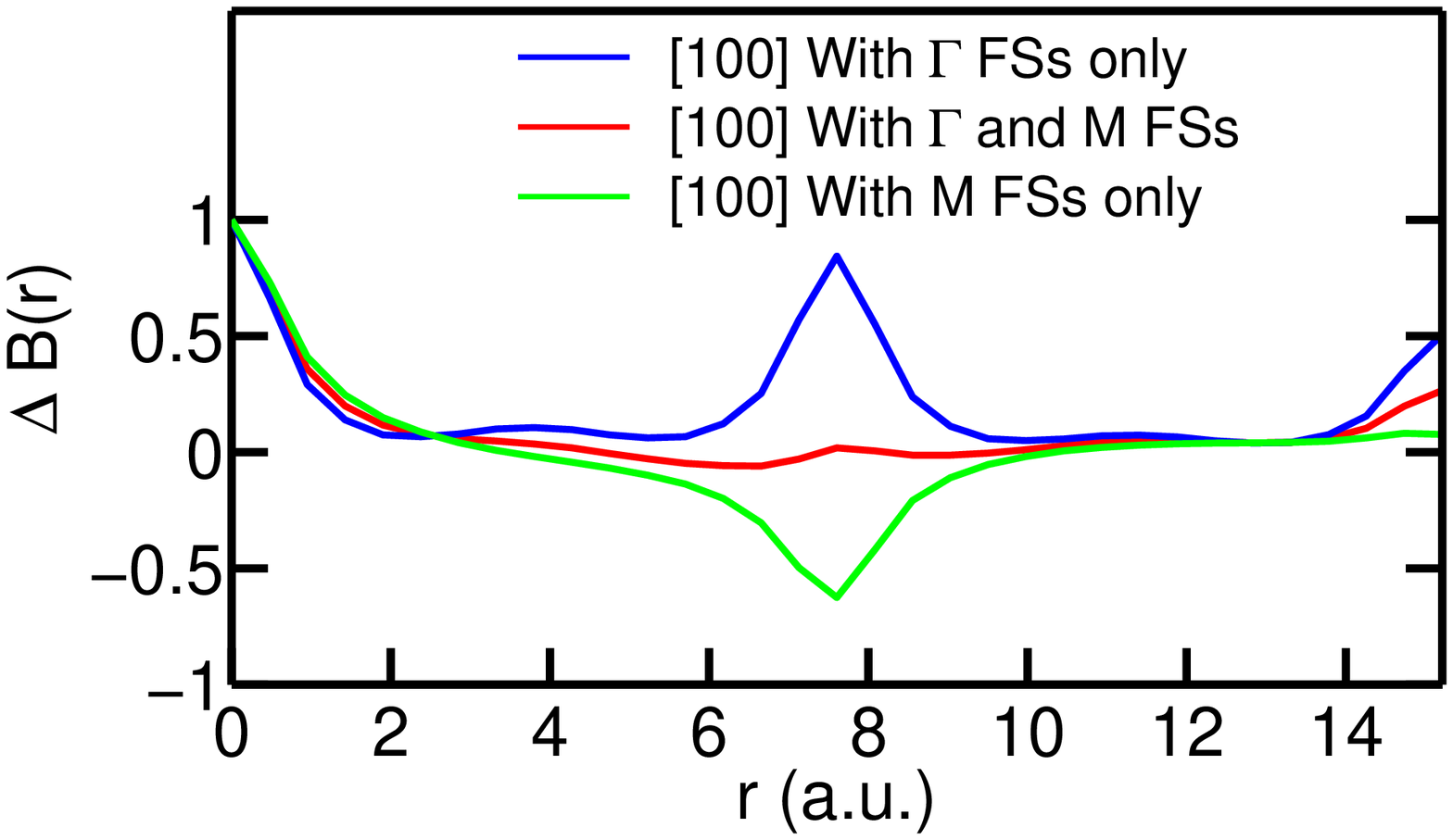}
\caption{(Color online)
Cut of $\Delta B$ along the [100] direction.
}
\label{cut}
\end{figure}

Interestingly, in Fig.~\ref{cut}, we show\cite{footnote2}
the contrasting behavior of the autocorrelation function
$\Delta B$ near the second neighbor Fe, at a
distance about 7.6 a.u. from the origin along the [100]-direction.
The minimum is produced by anti-bonding states belonging to
the M FSs while the maximum is the result of bonding states belonging to
the $\Gamma$ FSs. Hence with doping the feature should evolve from the
red curve when both FS are occupied to the green curve when the $\Gamma$ FS disappears.

Since this last analysis will need only one FT of the
experimental Compton profile along [100] it clearly provides particularly
robust FS information in real space. In contrast, the $\rho^{2D}$
reconstruction requires many more numerical manipulations \cite{matsumoto}
and therefore it can be less reliable. Our $\Delta B$ methodology
thus is a more reliable way to detect when the FS signal at $\Gamma$
vanishes.

In conclusion, our study predicts that high-resolution Compton scattering
spectra will yield signatures for the FS topology of Fe-based
superconductors. In particular, the $\Delta B$-map projected along the
[100] direction displays a remarkable signature of the FS evolution with
doping. Thus our method gives a robust way to establish the topology
rather than the precise shape of the iron pnictides FS.
These results indicate that Compton scattering can provide a
powerful new spectroscopic window for investigating FSs compatible with
the $s_\pm$ model for the description of the superconducting order
parameter in the Fe-based superconductors.

We are grateful to M. Lindroos, Y. Sakurai, Z. Hasan and T. Jarlborg
for important discussions.
This work is supported by the US Department of Energy, Office of
Science, Basic Energy Sciences contract DE-FG02-07ER46352, and
benefited from the allocation of supercomputer time at NERSC and
Northeastern University's Advanced Scientific Computation Center (ASCC).
It was also sponsored by the Stichting Nationale Computer Faciliteiten (NCF)
for the use of supercomputer facilities, with financial support from NWO
(Netherlands Organization for Scientific Research).


\begin{thebibliography}{99}

\bibitem{mazin}
I. I. Mazin \textit{et al.},
Phys. Rev. Lett. {\bf 101}, 057003 (2008).

\bibitem{mazin2}
I. I. Mazin and J. Schmalian, 
Physica C {\bf 469}, 614 (2009).

\bibitem{chubukov}
A. V. Chubukov, D. V. Efremov, and I. Eremin,
Phys. Rev. B {\bf 78}, 134512 (2008).

\bibitem{chubukov2}
A. V. Chubukov, I. Eremin and M. M. Korshunov,
Phys. Rev. B {\bf 79}, 220501(R) (2009).

\bibitem{norman}
M. R. Norman, Physics {\bf 1}, 21 (2008).

\bibitem{singh}
D. J. Singh and M.-H. Du,
Phys. Rev. Lett. {\bf 100}, 237003 (2008).

\bibitem{mazin3}
I. I. Mazin \textit{et al.},
Phys. Rev. B {\bf 78}, 085104 (2008).

\bibitem{haule}
K. Haule, J. H. Shim, and G. Kotliar,
Phys. Rev. Lett. {\bf 100}, 226402 (2008).

\bibitem{Terashima}
K. Terashima \textit{et al.},
Proc. Nat. Acad. Sci. {\bf 106}, 7330 (2009).

\bibitem{dhlu}
D. H. Lu \textit{et al.},
Physica C {\bf 469}, 452 (2009).

\bibitem{Sekiba}
Y. Sekiba \textit{et al.},
New J. Phys. {\bf 11}, 025020 (2009).

\bibitem{Sahrakorpi}
S. Sahrakorpi \textit{et al.},
Phys. Rev. Lett. {\bf 95}, 157601 (2005).

\bibitem{AIC}
A. I. Coldea \textit{et al.},
Phys. Rev. Lett. {\bf 101}, 216402 (2008).

\bibitem{jga}
J. G. Analytis \textit{et al.},
Phys. Rev. Lett. {\bf 103}, 076401 (2009).

\bibitem{footnote_posi}
Positron annihilation would provide another bulk-sensitive probe
of the Fermi surface. See, e.g., Ref. 16.

\bibitem{ft_posi1}
L. C. Smedskjaer \textit{et al.},
J. Phys. Chem. Solids {\bf 52}, 1541 (1991); P. E. Mijnarends {\it et
al.}, J. Physics: Conden. Matter 10, 10383 (1998).

\bibitem{Tanaka}
Y. Tanaka \textit{et al.},
Phys. Rev. B {\bf 63}, 045120 (2001).

\bibitem{Huotari}
S. Huotari \textit{et al.},
Phys. Rev. B {\bf 62}, 7956 (2000).

\bibitem{Stutz}
G. Stutz \textit{et al.},
Phys. Rev. B {\bf 60}, 7099 (1999).

\bibitem{cooper}
M.J. Cooper \textit{et al.} (editors),
\textit{X-Ray Compton Scattering}, Oxford University Press, Oxford (2004).

\bibitem{Kaplan}
I. G. Kaplan, B. Barbiellini and A. Bansil,
Phys. Rev. B {\bf 68}, 235104 (2003).

\bibitem{matsumoto}
I. Matsumoto \textit{et al.},
Phys. Rev. B {\bf 64}, 045121 (2001).

\bibitem{ABkkr}
A. Bansil \textit{et al.},
Phys. Rev. B {\bf 60}, 13396 (1999).

\bibitem{Qiu2008}
Y. Qiu \textit{et al.},
Phys. Rev. B {\bf 78}, 052508 (2008).

\bibitem{footnote_sophi}
A more sophisticated treatment using KKR-CPA or other approaches
[see, e.g., Refs. 26-29] was not undertaken.

\bibitem{ft_sophi1}

A. Bansil, Zeitschrift Naturforschung A {\bf 48}, 165 (1993); A. Bansil,
Phys. Rev. B20, 4035 (1979).

\bibitem{ft_sophi2}
L. Schwartz and A. Bansil,
Phys. Rev. B {\bf 10}, 3261 (1974).

\bibitem{ft_sophi3}
S. N. Khanna \textit{et al.},
Solid State Commun. {\bf 55}, 223 (1985).

\bibitem{ft_sophi4}
H. Lin \textit{et al.},
Phys. Rev. Lett. {\bf 96}, 097001 (2006).

\bibitem{bba01}
B. Barbiellini, A. Bansil,
J. Phys. Chem. Solids {\bf 62}, 2181 (2001).

\bibitem{lcw1973}
D.G. Lock, V.H.C. Crisp, and R.N. West,
J. Phys. F: Met. Phys. {\bf 3}, 561 (1973).

\bibitem{weiku}
C.-C. Lee, W.-G. Yin and W. Ku, 
Phys. Rev. Lett. {\bf 103}, 267001 (2009).

\bibitem{footnote1}
The exact value of
$x_2$ is sensitive to the assumption of
rigid band filling and to the exchange-correlation
functional used (See Refs. 7 and 8).

\bibitem{footnote_filter}
This filtering procedure can also be applied to experimental Compton data.

\bibitem{footnote2}
The cut of $\Delta B$ with $\Gamma$ FS is derived from
$\Delta B|_{x_1}^{x_2}(x,y)$ for $x_2=0.15$ and $x_1=0.10$,
while the cut of $\Delta B$ without $\Gamma$ FS is obtained from
$\Delta B|_{x_1}^{x_2}(x,y)$ for $x_2=0.15$ and $x_1=0.10$,
after artificially removing all the Bloch states around $\Gamma$.

\end{thebibliography}
\end{document}